\documentclass{article}

\usepackage{arxiv}

\usepackage[utf8]{inputenc} 
\usepackage[T1]{fontenc}    
\usepackage{hyperref}       
\usepackage{url}            
\usepackage{booktabs}       
\usepackage{amsfonts}       
\usepackage{nicefrac}       
\usepackage{microtype}      
\usepackage{lipsum}
\usepackage{amsmath}
\usepackage{algorithmic}
\usepackage{graphicx}
\usepackage{textcomp}
\usepackage{float}
\usepackage{lineno}
\usepackage{booktabs}
\usepackage{multirow}
\usepackage{siunitx,subcaption}

\title{Decision and Feature Level Fusion of Deep Features Extracted from Public COVID-19 Data-sets}

\author{
  Hamza Osman Ilhan \\
  Department of Computer Science\\
  Yildiz Technical University\\
  Istanbul, Turkey \\
  \texttt{hoilhan@yildiz.edu.tr} \\
   \And
 Gorkem Serbes\\
  Department of Biomedical Engineering\\
  Yildiz Technical University\\
  Istanbul, Turkey \\
  \texttt{gserbes@yildiz.edu.tr} \\
   
   \And
   
  Nizamettin Aydin \\
  Department of Computer Science\\
  Yildiz Technical University\\
  Istanbul, Turkey \\
  \texttt{naydin@yildiz.edu.tr} \\
}

\begin{document}
\maketitle

\begin{abstract}
{The Coronavirus disease (COVID-19), which is an infectious pulmonary disorder, has affected millions of people and has been declared as a global pandemic by the WHO. Due to highly contagious nature of COVID-19 and its high possibility of causing severe conditions in the patients, the development of rapid and accurate diagnostic tools have gained importance. The real-time reverse transcription-polymerize chain reaction (RT-PCR) is used to detect the presence of Coronavirus RNA by using the mucus and saliva mixture samples taken by the nasopharyngeal swab technique. But, RT-PCR suffers from having low-sensitivity especially in the early stage. Therefore, the usage of chest radiography has been increasing in the early diagnosis of COVID-19 due to its fast imaging speed, significantly low cost and low dosage exposure of radiation. In our study, a computer-aided diagnosis system for X-ray images based on convolutional neural networks (CNNs) and ensemble learning idea, which can be used by radiologists as a supporting tool in COVID-19 detection, has been proposed. Deep feature sets extracted by using seven CNN architectures were concatenated for feature level fusion and fed to multiple classifiers in terms of decision level fusion idea with the aim of discriminating COVID-19, pneumonia and no-finding classes. In the decision level fusion idea, a majority voting scheme was applied to the resultant decisions of classifiers. The obtained accuracy values and confusion matrix based evaluation criteria were presented for three progressively created data-sets. The aspects of the proposed method that are superior to existing COVID-19 detection studies have been discussed and the fusion performance of proposed approach was validated visually by using Class Activation Mapping technique. The experimental results show that the proposed approach has attained high COVID-19 detection performance that was proven by its comparable accuracy and superior precision/recall values with the existing studies.}
\end{abstract}


\section{Introduction}
\label{sec:introduction}
The coronavirus disease 2019 (COVID-19) is a respiratory disorder, which may have varying severity respiratory symptoms from the common cold to fatal pneumonia. COVID-19 is caused by a novel coronavirus known as the severe acute respiratory syndrome coronavirus 2 (SARS-CoV2). The first cases of the COVID-19, which were reported as pneumonia patients with unknown cause, were originated from Wuhan, Hubei province of China on December, 2019. The COVID-19 has spread to most of the provinces of China in 30 days \cite{wu2020characteristics} and the condition has been declared as a global pandemic by the World Health Organization (WHO) in a very short time \cite{sohrabi2020world, lai2020severe}. It was thought that the epicenter of the outbreak was a wholesale market selling seafood and other exotic animals, including bats, marmots, and snakes \cite{sohrabi2020world}. SARS-CoV2 has very high contagious nature with a 1-14 days long incubation period. Some of the carriers may not show any symptoms while a significant amount of the patients may have minor symptoms such as dry-cough, sore throat, headache, fatigue, and sputum production. However, the virus can be fatal if the immune system of the patient is weak \cite{razai2020coronavirus}. The conditions seen in the severe and critical patients may be the pneumonia, acute respiratory distress syndrome, pulmonary edema or multiple organ failure \cite{chen2020epidemiological,chung2020ct}. In \cite{clerkin2020covid}, it was stated that approximately 14\% of the COVID-19 patients have experienced severe conditions such as the dyspnea,  while 5\% of the patients were in critical condition including respiratory failure, septic shock, or multiple organ dysfunction. Early diagnosis of the COVID-19 and the application of successful treatment is the key factor to reduce the complication and mortality in patients having underlying medical conditions such as hypertension, diabetes, cardiovascular disease and asthma \cite{fang2020patients,li2020prevalence,clerkin2020covid,hegde2020does}.

Another important factor related with the COVID-19 is the transmission mechanism of the SARS-CoV2. The primary propagation mechanism of the SARS-CoV2 has been identified as the spread of respiratory droplets through sneezing and coughing, which have the potential to cover a distance up to 1.8 meters \cite{chamola2020comprehensive}. This highly contagious nature of the SARS-CoV2 puts any person, who has a close contact history with the patient, in a very high risk. Although, the primary source of the SARS-CoV2 transmission has been identified as the symptomatic people, asymptomatic people can also have a possibility to be a risk factor \cite{chamola2020comprehensive}. The higher risk of getting severe COVID-19 disease for the patients having existing medical conditions and being over age 60 years, and the high potential of fast propagation risk of COVID-19 results in a significant need for the fast and accurate diagnosis tools.

As the most common test technique to diagnose COVID-19, the real-time reverse transcription-polymerase chain reaction (RT-PCR) is used to detect the presence of viral RNA. In this method, a sample including a mixture of mucus and saliva is taken by using the nasopharyngeal swab technique for being assessed for virus existence. Besides, the specimen is recommended to be taken form the lower respiratory tract of the patient when severe respiratory ailments are seen \cite{pfefferle2020evaluation}. However, the RT-PCR suffers from having low-sensitivity especially in the early stage \cite{li2020coronavirus,fang2020sensitivity} and it was mentioned in \cite{wang2020temporal} that the chest radiography has performed very well in the early diagnosis of COVID-19. Therefore, it is believed that complementing the nucleic acid testing with chest radiography based diagnosis has promising potential in the early detection of COVID-19 \cite{kanne2020essentials}. 

Regarding the chest radiography techniques, X-rays and Computer tomography (CT) scans are the most commonly used imaging methods to diagnose the thoracic abnormalities. Although the CT scan can provide finer details of the 3D anatomy of human body, X-rays are more convenient to differentiate between viral and non-viral pneumonia due to its fast imaging speed, significantly low cost and low dosage exposing of radiation \cite{self2013high}. Furthermore, in \cite{jacobi2020portable}, the most common manifestations and patterns of lung abnormality on portable chest radiography (CXR) in COVID-19 were described and it was mentioned that the CXR will likely be the most commonly utilized method for diagnosis and follow up of COVID-19 because of the infection control issues related to patient transport to CT suites, the problems experienced in CT room decontamination, and lack of CT availability in parts of the world. In \cite{borghesi2020covid}, an experimental CXR scoring system, which was tested on hospitalized patients with COVID-19 pneumonia, was presented to quantify and monitor the severity and progression of disease. The authors found that the inter-observer agreement of the developed system was very good and the CXR based scoring is a promising tool for predicting mortality in hospitalized patients with SARS-CoV2 infection. In the light of the advantages of X-ray imaging over CT scan in the diagnosis and monitoring of COVID-19, we focus on developing a X-ray imaging based automated system which has the ability of differentiating viral
pneumonia (COVID-19) from non-viral pneumonia and normal controls (No findings).

Computer-aided diagnosis (CAD) has been successfully used as a supporting tool for the diagnosis process of radiologists since 1980s \cite{doi2007computer}. The CAD systems are mostly developed as a complementary decision making approach to the diagnosis of physicians due to their advantages such as being reproducible and having the ability of detecting subtle changes that cannot be observed by the visual inspection. With respect to the usage of X-ray imaging based CAD systems in the diagnosis of thoracic diseases, the recent advances in deep learning have led to breakthrough improvements in the discrimination of viral and non-viral pneumonia. In \cite{kermany2018identifying}, a diagnostic tool, which is based on a deep-learning framework for diagnosis of pediatric pneumonia using chest X-ray images, was proposed. In \cite{rajaraman2018visualization}, the performance of customized convolutional neural networks (CNNs) to differentiate between bacterial and viral pneumonia types in pediatric CXRs was presented. Additionally, various deep learning approaches were successfully employed to diagnose pneumonia and other pathologies in \cite{baltruschat2019comparison,yadav2019deep,jaiswal2019identifying}. In order to detect COVID-19 samples by using X-rays, a deep learning architecture, which employs depthwise convolutions with varying dilation rates to incorporate local and global features extracted from diversified receptive fields, was presented in \cite{mahmud2020covxnet}. In \cite{tougaccar2020covid}, various deep learning models were utilized for feature extraction and the obtained feature sets were processed using the Social Mimic optimization method. Later, the modified deep features were given to support vector machines (SVMs) with the aim of COVID-19 detection. In \cite{rahimzadeh2020modified}, a concatenated neural network, which is based on Xception and ResNet50V2 networks for classifying the chest X-ray images into three categories of normal, pneumonia, and COVID-19, was presented in an unbalanced data-set configuration. In \cite{oh2020deep}, a patch-based CNN approach with a relatively small number of trainable parameters was given for COVID-19 diagnosis. In this method, random patches were cropped from the X-ray images and the final classification result was obtained by majority voting from inference results at multiple patch locations. In \cite{elasnaoui2020using}, a comparative individual analysis of the recent deep learning models including VGG16, VGG19, DenseNet201, InceptionResNetV2, InceptionV3, Resnet50, and MobileNetV2 was presented in the detection and classification of COVID-19. An Auxiliary Classifier Generative Adversarial Network based model was employed in \cite{waheed2020covidgan} for generating synthetic chest X-ray CXR images to avoid overfitting and increase the generalization capability of employed CNNs. In \cite{ozturk2020automated}, an end-to-end deep learning architecture, which was an enhanced version of the Darknet-19 model, was employed for the multi-class classification (COVID vs. No-Findings vs. Pneumonia).

Although previous studies have shed some lights on the deep learning-based diagnosis by using X-ray images and significant improvement has been obtained, none of the previous works have been able to propose a complete solution to the COVID-19 detection problem. Additionally, the COVID-19 outbreak is recent and the content of the public X-ray imaging databases is still progressing. Due to this gradual increase in the number of COVID-19 images in the public databases, a need of developing new algorithms, which have generalization capability for new COVID-19 samples, have been raised. In this study, we propose a deep features based ensemble learning model, which uses feature and decision level fusion, in order to satisfy the aforementioned needs in COVID-19 diagnosis. The main contributions of this study are summarized as follows: i) The proposed learning model was applied to progressively created three public COVID-19 databases in order to measure its generalization capability and reduce the biasing effect that can occur in unbalanced databases. ii) The individual performance of seven powerful deep learning architectures including the Mobilenet, VGG16, ResNet50, ResNet101, NasNet, InceptionV3 and Xception were presented. iii) The same seven deep learning models were employed as feature extractors and the obtained individual deep features were fed to non-linear kernel SVMs with the aim of COVID-19 detection. iv) The extracted deep features by using individual CNNs were concatenated to form a single feature vector (feature level fusion) which was subsequently given to classifiers. v) The decisions of the individual classifiers were combined by employing the majority voting schema (decision level fusion). vi) The experimental results have demonstrated the effectiveness and robustness of the proposed ensemble approach in epidemic screening by reaching high general accuracy values accompanied with high COVID-19 F1-scores, precision and recall values. The rest of the study is organized as follows; Section \ref{MaterialsAndMthods} introduces materials and methods. Section \ref{ExperimentalResults} presents the experimental results and finally, Section \ref{DiscussionandConclusion} presents the discussion and conclusion.

\section{Materials and Methods}
\label{MaterialsAndMthods}
In this study, an ensemble of CNNs with the aid of decision and feature level fusion idea was proposed to solve the classification problem in X-ray images for COVID-19, No-Findings and Pneumonia classes. For doing that three public X-ray datasets were employed in the experiments and the generalization capability of the proposed approach has been proven. In the ensemble of CNNs, transfer learning layout of seven deep convolutional neural network (CNN) models, which were initially pre-trained by a huge image collection repository, the ImageNet, were utilized. The employed deep networks, whose individual classification performance were also given, were the MobilenetV2, VGG16, ResNet50, ResNet101, NasNet, InceptionV3 and Xception. In addition, the same seven deep networks were also employed as deep feature extractors and the obtained deep features were fused and the resultant concatenated feature vector was fed to non-linear kernel based SVMs to increase the discrimination performance. 

\subsection{Dataset Information}
In our study, three databases were constructed in a progressive way to measure the classification performance and generalization ability of the proposed approach by using the combinations of three publicly available data-sets. Firstly, the data-set that has been already used in \cite{ozturk2020automated} was employed as the baseline reference database and it is named as DB1. DB1 consists of 126 COVID-19 images, 500 pneumonia images and 500 normal (no-finding) images. The COVID-19 images of DB1 were taken from a public data-set, which is constantly updated by researchers \cite{cohen2020covid}. The remaining 1000 non-COVID X-ray images were taken from the public ChestX-ray8 dataset \cite{wang2017chestx} and the DB1 was finalized with 1126 X-ray images. Secondly, at the date of this study, 80 new COVID-19 samples, which have been appended to DB1 by researchers after the publication of \cite{ozturk2020automated}, were added to DB1 to be able to compare our study with other state-of-art findings. This new database, which contains 1206 X-ray images in total, is named as DB2. Lastly, 113 new COVID-19 samples obtained from different domain were added to DB2 to be able to create a more balanced data-set that would be more convenient to measure performance of the proposed method. The new  113 COVID-19 samples were taken from \cite{Agchung} resulting in the DB3, which  contains 1319 X-ray images in total. A short summary of the constructed data-sets for the proposed study is given in Table \ref{dataDist}.

\begin{table}[!ht]
\centering
\caption{The Image Distributions over Classes in Tested Datasets}
\label{table_1}
\begin{tabular}{lccc}
Labels & DB1 \cite{ozturk2020automated} &  DB2 \cite{ozturk2020automated}+\cite{cohen2020covid} &  DB3 \cite{ozturk2020automated}+\cite{cohen2020covid}+\cite{Agchung} \\ \hline
COVID-19   & 126      &    206     &      319    \\ 
No Findings   & 500    & 500    & 500  \\ 
Pneumonia   & 500   & 500  & 500         \\  \hline
Total   & 1126  & 1206       & 1319 \\ \hline  
\label{dataDist}
\end{tabular}
\end{table}

\subsection{Employed Deep Learning Architectures}

The traditional machine learning approaches, which consist of sequential sub-steps such as pre-processing, feature extraction, feature reduction/selection and classification, require domain specific expertise in order to obtain satisfactory performance in medical image analysis. The spatial and frequency domain features are the most popular approaches to obtain discriminating information from the raw images. For example, the Scale-Invariant Feature Transform (SIFT) and Maximally Stable Extreme Regions (MSER) methods are used in literature \cite{ilhan2020fully,ilhan2020automated} as the spatial domain interest point extraction techniques and the interest points based features are employed in traditional learning models subsequently. Regarding the frequency domain feature extractors like short time Fourier Transform (STFT) and wavelet transform (WT), the parameter selection procedure makes them hard to implement and dependent to user experience. For example, the window-type, window-length and overlapping ratio must be chosen in an optimum way for STFT, while the mother wavelet type and decomposition level must be correctly tuned in WT to obtain a representative distribution of time-scale atoms \cite{serbes2011effect,ilhan2018dual}. On the other hand, even if the training processing times of deep learners are relatively long, they are implemented in end-to-end architectures which have no need or having minimum need for extra pre-processing steps and optimum tuning of feature extractor parameters. Additionally, deep neural networks (DNNs), which gained importance in machine learning and pattern recognition in recent years, have the ability to learn high order robust features from raw data \cite{ravi2016deep,afshar2019handcrafted}. In contrast, traditional machine learning methods are still highly error prone and inaccurate to be used in a sensitive decision making process. Therefore, in order to benefit from the aforementioned superiorities of deep learners, seven CNN models including the MobileNetV2, VGG16, ResNet50, ResNet101, NasNet, InceptionV3 and Xception, have been applied to three public databases with the aim of three-class (COVID, No-Findings, Pneumonia) discrimination of X-ray images in the proposed study.

\subsubsection{MobileNetV2}
Although higher accuracy values can be achieved by using deeper and larger networks, these networks do not ensure efficiency in terms of size and speed, making them inconvenient for mobile applications. However, the fast and accurate diagnosis of COVID-19 is critical in the current pandemic condition causing the small mobile deep learning solutions more preferable. The MobileNetV2 \cite{sandler2018mobilenetv2}, as an improvement of MobileNetV1, can be a powerful and versatile solution for mobile dignosis of COVID-19 due to its high performance proven in various application areas including medieval writer identification \cite{cilia2020end}, detecting underwater live crabs \cite{cao2020real},  real-time crowd counting \cite{gao2019mobilecount} and 
remote wave gauging \cite{buscombe2020optical}. The main characteristic of MobileNetV2 relies on the usage of depthwise separable convolutions in which two 1D convolutions with two kernels are used instead of employing 2D convolution with a single kernel. As a result, the training phase can be carried out by using fewer parameters and less memory that results in a small and efficient model.     

\subsubsection{VGG16}
The VGG16 \cite{simonyan2014very} is a pre-trained very large CNN that was invented by VGG (Visual Geometry Group) from University of Oxford. The VGG16 was the 1st runner-up of the ILSVR (ImageNet Large Scale Visual Recognition Competition) 2014 in the classification task. The VGG16 architecture uses simple 3×3 size kernels in convolutional layers and combine them in a sequence to emulate the effect of larger receptive fields. The implemented VGG16 architecture is composed of 13 convolutional layers followed by 3 fully connected layers. Despite the simplicity of the VGG16 architecture, its memory usage and computational cost is dramatically high due to the exponentially increasing kernels. 

\subsubsection{ResNet50 and ResNet101}
The ResNet deep learning models \cite{he2016deep}, which have introduced the “skip connections” concept, are the sub-classes of CNNs. In ResNets, some of the convolutional layers are bypassed (the concept of “skip connections”) at a time and the batch normalization is applied along with non-linearities (ReLU) \cite{pacheco2020impact}. In ResNet architectures, the "skip connections" enables to train much deeper networks and they give the network the option to simply copy the activations from ResNet block to ResNet block, preserving information as data goes through the layers \cite{lundervold2019overview}. The two versions of ResNet family, the ResNet50 and ResNet101 having 49 and 100 convolutional layers respectively, were employed in the current proposed COVID-19 diagnosis approach as a classifier and deep feature extractor. 

\subsubsection{NasNet}
As a relatively recent network, the NASNet \cite{zoph2018learning}, whose CNN architecture was designed by another neural network, outperformed the previous state-of-the-art on the ILSVRC 2012 dataset. The NASNet architecture was created by use of the Neural Architecture Search (NAS) framework providing an algorithm for finding optimal neural network architectures \cite{cogan2019mapgi}. In this algorithm, a controller recurrent neural network creates architectures aimed to perform at a specific level for a particular task, and by trial and error learns to propose better and better models \cite{lundervold2019overview}.

\subsubsection{InceptionV3}
In the InceptionV3 \cite{szegedy2016rethinking}, the inception modules, which are repeatedly stacked together to form a large network, are employed as an alternative to sequentially ordered convolution layers. In the inception modules, an asymmetric convolution structure is obtained by using multiple filters of various sizes resulting in more and more abundant spatial features with increased diversity. The usage of inception modules not only cause significant decrements in the number of parameters, it also increases the recognition ability of the network by using multiple scale features \cite{zhuang2019detection}.  

\subsubsection{Xception}
As an improved version of inception architecture, the Xception \cite{chollet2017xception} algorithm uses depthwise separable convolutions which enables more efficient use of model parameters. In the Xception, the standard inception modules are replaced with the depthwise separable convolutions (enhanced inception modules) that use the depth dimension (the number of channels) as well as the spatial information. The enhanced inception modules result in stronger features including the depth information.

\subsection{Transfer Learning}
In the context of deep learning based classification, transfer learning involves training a deep-net on a labelled training dataset (consisting of high number of samples) in order to circumvent the obstacle of insufficient number of training samples. For the analysis of medical images, the weights of the deep-net learned during the training of a CNN on a main dataset (for example ImageNet \cite{russakovsky2015imagenet}) are transferred to a second CNN, which is then re-trained on labelled samples of desired medical data set using pre-learned weights. The final training phase is named as "fine tuning"; in which the certain layers of pre-trained net can be frozen (the weights of these layers stay fixed) while the remaining layers can be fine-tuned to suit the classification problem, except the last fully connected layer.

In our study, the employed CNNs were applied to COVID-19 data-sets by using the Transfer Learning strategy in the light of literature findings. In \cite{azizpour2015factors}, it was mentioned that the performance of knowledge transfer depends on the dissimilarity between the database on which a CNN is trained and the database to which the knowledge is to be transferred. The distance between the natural image databases, that are employed for knowledge transfer, and COVID-19 data-sets is considerable. However, recent studies show that there is a potential for having benefit from knowledge transfer in medical data sets. For instance, in \cite{bar2015deep}, a pre-trained CNN was employed as a feature extractor with the aim of chest pathology identification. In \cite{van2015off}, pre-trained CNN based features have shown improved performance as they were fused with traditional handcrafted features in a nodule detection system. In addition to their feature extractor usage, the knowledge transferred CNNs can also be employed as the main classification framework with fine-tuning. For example, in \cite{tajbakhsh2016convolutional}, it was shown that the fine-tuned CNNs have performed similarly or better than the CNNs trained from scratch. In this study, pre-trained weights from \cite{krizhevsky2012imagenet} were transferred in either a shallow tuning or deep tuning strategy in which the weights of few layers for the former and many layers for the latter were trained. The results highlighted that medical image analysis requires deep tuning of more layers in contrast to many other computer vision tasks. In another study, it was demonstrated that fine-tuning of pre-trained networks worked better compared to nets trained from scratch in the analysis of skin lesions \cite{menegola2016towards}. Additionally, in \cite{shin2016deep} knowledge transfer from natural images was applied in thoraco-abdominal lymph node detection and interstitial lung disease classification resulting in higher performance than training the CNNs from scratch. Similarly, in \cite{chen2015standard}, transfer learning strategy was employed to identify the fetal abdominal standard plane and the approach revealed improved capability of the algorithm to encode the complicated appearance of the abdominal plane. In our study, due to the aforementioned superiority of fine-tuning strategy, seven CNNs,  which have already been trained by natural image database (ImageNet), were fine-tuned to extract deep features by using the X-ray samples. Later, these deep features were employed in the classification of chest X-ray images with individual and ensemble learning models.

\subsection{Decision and Feature Level Fusion}
In a pattern recognition system, the ultimate goal is the design of best possible classification model for a specific problem such as the COVID-19 detection by using X-ray images. Traditionally, various classification models that have different theories and methodologies are applied to a specific pattern recognition problem, and the best model in terms of performance metrics is chosen. However, it was observed that the sets of patterns misclassified by the various classifiers would not necessarily overlap, even if one of the models has yielded the best accuracy. Hence, different classifiers may be harnessed to improve the overall performance by using their possible complementary information about the patterns to be classified, when they are used in an ensemble scheme \cite{kittler1998combining}. This type of ensemble learning scheme is called decision level fusion based learning, in which the individual decisions of different models are combined to derive a consensus decision instead of relying on a single decision-making model. The hard-level combination uses the individual outputs of each classifier after they are binarized by applying a threshold to the classifier output probabilities (estimates of a posteriori probability of the class) to map them into class labels. As a member of hard-level combination, the majority voting strategy simply counts the votes received from each classifier and the class that has the largest number of votes is selected as the consensus decision.

As an additional fusion strategy, the feature level fusion, in which various sets of features obtained by different feature extractors are combined, has high potential to achieve better classification performance \cite{li2004fusing,ulukaya2017overcomplete,sakar2019comparative,gunatilaka2001feature}. Feature level fusion generally consists of the concatenation of various normalized feature subsets resulting in a single feature vector forming a complete representation of different views (deep features obtained by using various CNNs). Regarding the CNNs based feature level fusion studies, even if the various CNN models are based on different configurations (or architectures), the fine-tuning of these CNN models by using the same target database (COVID-19 database in our study) consisting of concatenated feature vectors, can provide complementary information \cite{raja2017transferable,du2018feature}.  

\subsection{Proposed Deep Features Based Ensemble Model}

\begin{figure*}[t]
    \centering
    \includegraphics[width=\textwidth]{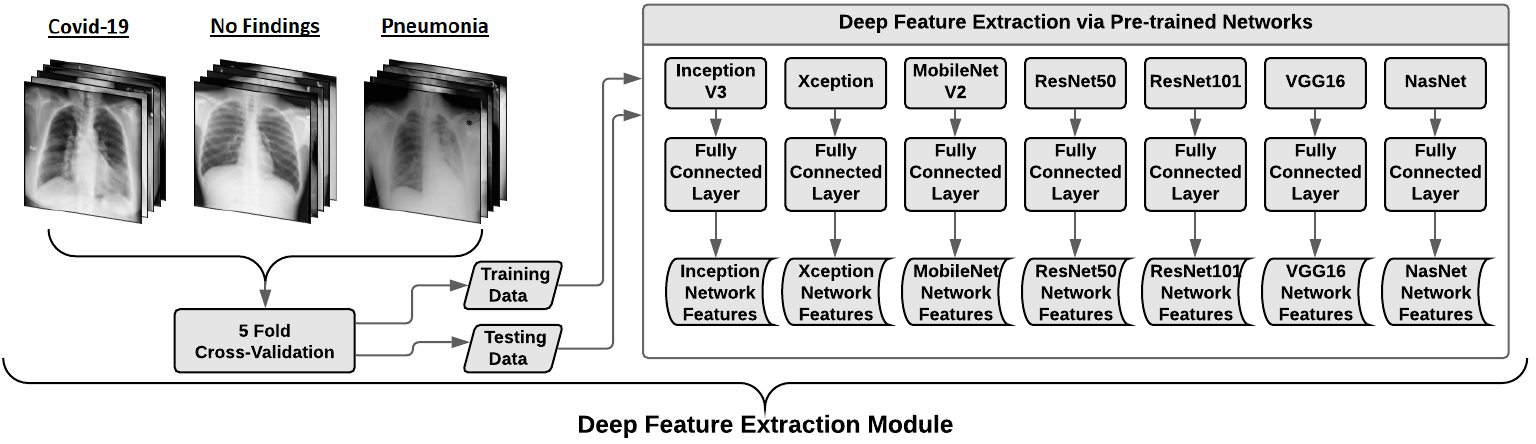}
    \caption{The deep feature extraction module}
    \label{fig:deep_feature_extractor}
\end{figure*}
\begin{figure*}[t]
    \centering
    \includegraphics[width=\textwidth]{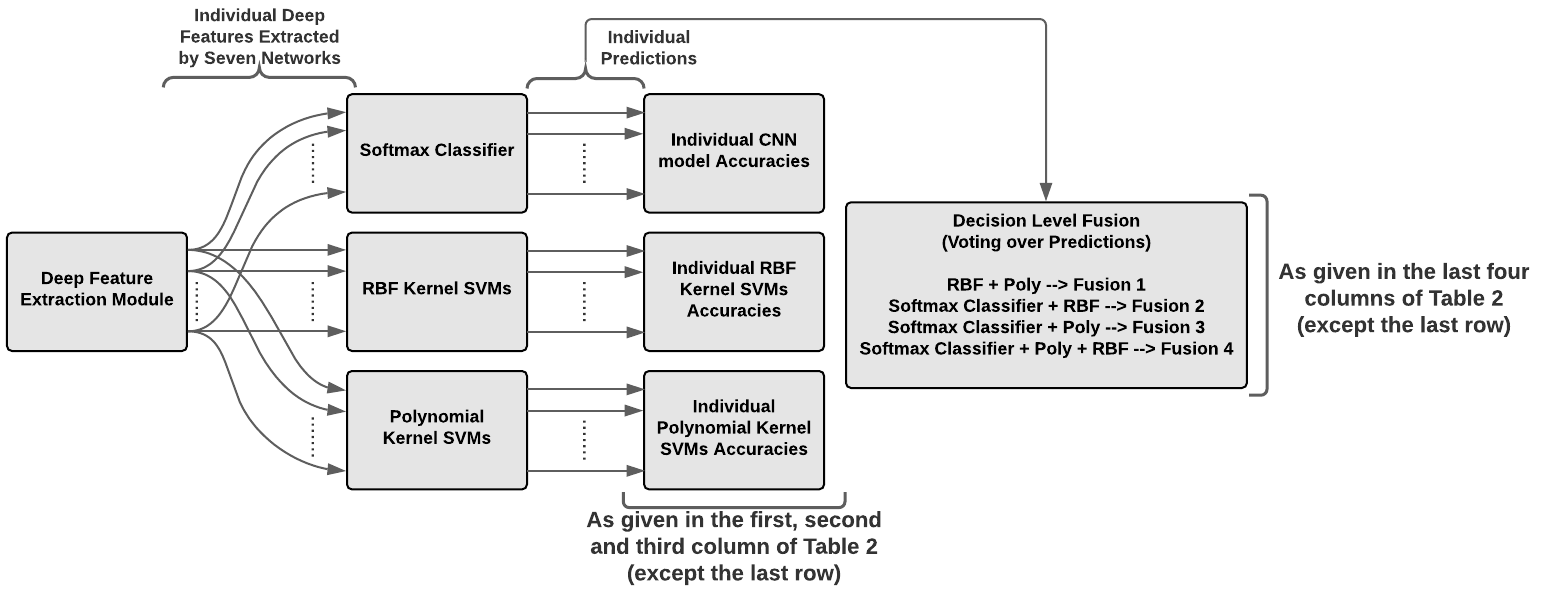}
    \caption{The multistage learning approach and decision level fusion of individual classifiers}
    \label{fig:individual_learners_and_fusion_figure}
\end{figure*}

In this study, seven CNN models (the MobilenetV2, VGG16, ResNet50, ResNet101, NasNet, InceptionV3, and Xception) have been used as the main structure of proposed framework. During the development of proposed method, firstly, these seven CNN models have been employed as deep feature extractors as depicted in Figure \ref{fig:deep_feature_extractor}. As seen in Figure \ref{fig:deep_feature_extractor}, the three databases were fed to the individual CNNs, which have already been pre-trained by using ImageNet \cite{russakovsky2015imagenet}, with the aim of network specific deep feature extraction by using a 5-fold cross-validation scheme. The optimum hyperparameters were chosen by employing a batch-size, epoch, and learning rate analysis that was based on trial and error strategy. Accordingly, the number of training epochs was chosen as 50, while a batch-size of 16 was employed. The learning rate that controls the speed of convergence was set to 0.0001, when Stochastic Gradient Descent with momentum was used as the optimization technique. Subsequent to the deep feature extraction phase, the obtained deep features were fed to a softmax classifier satisfying the end-to-end learning scheme of classical deep learning. The classical softmax layer of CNNs, which is the generalization of logistic sigmoid function with the ability of mapping deep-features onto probability values used as outputs in discrimination problems having three or more classes, is named as "softmax classifier" in our study. The softmax classifier \cite{khamparia2019kdsae,sharif2020active,liao2015image} is employed to measure the discriminating power of deep features obtained from the individual CNNs and also for the concatenated deep feature vector. The predictions of the individual CNNs were obtained as shown in the second-column/upper-row of Figure \ref{fig:individual_learners_and_fusion_figure}. 

In \cite{huang2006large} and \cite{tang2013deep}, it was mentioned that the CNNs, which are very good at learning invariant features, may show lower performance than the SVMs in classification. On the other hand, the SVMs are very successful at producing optimal decision surfaces from well behaved feature vectors, while having difficulties to represent the variances occurred in image features. Regarding the chest X-ray images used in our study, the areas that characterize the lung consolidation pattern may be located in various parts of the lung with changing size resulting in significant variances. Therefore, in addition to individual CNN based learning, a multistage model, in which the CNNs are employed to extract deep features that have potential to detect and recognize lung consolidation patterns, and non-linear SVMs that are trained by feeding the deep features learned by the CNNs, was presented and its performance was validated by using three databases. This multistage learning approach that uses CNNs and SVMs in a cascade connection has been successfully employed in various areas with the aim of classification performance improvement \cite{liang2019rice,su2020fusing,fuhad2020deep}. In this configuration, fully-connected activations of each CNN have been employed as feature extractors (given in Figure \ref{fig:deep_feature_extractor}) and the obtained deep feature vectors were fed to classifiers in a 5-fold validation scheme. Additionally, with the aim of performance improvement, the predictions obtained from classical end-to-end CNN learning (deep features that are fed to sotmax classifier) and kernel based SVMs (deep features that are fed to SVMs) were fused by using the voting approach in accordance with the combinations given in Table \ref{All_Results_Table}. The SVM based learning configuration that uses the deep features and the applied voting strategy was presented in Figure \ref{fig:individual_learners_and_fusion_figure}.

\begin{figure*}[t]
    \centering
    \includegraphics[width=\textwidth]{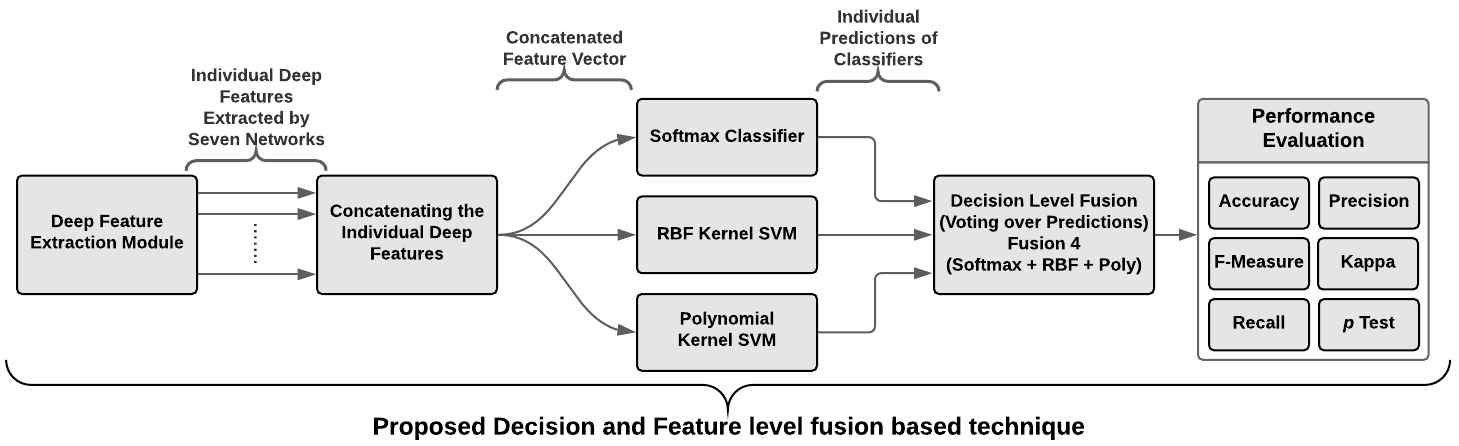}
    \caption{The Flowchart of the proposed method employing feature and decision level fusion.}
    \label{fig:proposed_method}
\end{figure*}

Regarding the feature level fusion phase; the deep features extracted by each employed CNNs were concatenated into a single fused feature vector. Subsequently, the fused feature vector was fed to the softmax classifier and also to the non-linear SVMs. After this, the individual predictions of the softmax classifier and SVMs were obtained as depicted in Figure \ref{fig:proposed_method}. Furthermore, to benefit from the possible complementary behaviour of the learning models, the obtained individual decisions were fused by using the majority voting and this final approach, which gives the best performance, was chosen as our proposed method. The detailed flowchart of proposed method including the deep feature extraction module, feature level fusion, multistage learning and decision level fusion can be seen in Figure \ref{fig:proposed_method}.

\section{Experimental Results}
\label{ExperimentalResults}
The individual performance of the employed seven CNN models plus the results of concatenated feature vector can be seen in the first column of Table \ref{All_Results_Table} in terms of the accuracy metric (each presented accuracy value was calculated as the mean of 5-folds and the standard deviation of these 5-folds were also represented for clarification). The highest accuracy values were obtained as 87.6\%, 85.7\% and 85.7\% for the DB1, DB2 and DB3 by employing the InceptionV3, ResNet50 and VGG16 respectively. In contrast, when the poorest individual performances are investigated, it is seen that the MobileNetV2 had the worst accuracy value as 84.2\% for DB1, while the NasNet has ended up with the accuracy values as 83.3\% and 84.1\% for DB2 and DB3 respectively. The second and third columns of Table \ref{All_Results_Table} show the accuracy values obtained by multistage learning scheme, which uses non-linear SVM kernels, for the individual deep feature sets and also for the concatenated feature vector as given in bottom row group. As seen in column 2, the highest accuracy value was obtained by using the RBF kernel as 87.6\% for the DB1 with no increment compared to softmax classifier. On the other hand, the RBF kernel based SVM learning, which were fed by VGG16 deep features, has slightly increased best accuracy value to 85.9\% for DB2, while the  InceptionV3 has reached to 86.2\% for DB3 by using RBF kernel based multistage approach. In addition, the columns 4, 5, 6 and 7 indicate the accuracy values obtained by using the decision level fusion strategy composed of the combinations of softmax classifier, radial basis function (RBF) and polynomial kernel based SVMs as highlighted in the Table \ref{All_Results_Table}.  

\begin{table}[!ht]
\centering
\scriptsize
\caption{The detailed presentation of accuracy values obtained from applied individual and ensemble learning scenarios for three data-sets. The standard deviations of accuracy values obtained from 5-folds are presented in parentheses.}
\resizebox{\textwidth}{!}{\begin{tabular}{|c|cl|ccc|cccc|}
\cline{1-10}
\multicolumn{3}{|c|}{\multirow{3}{*}{Accuracy (Standard Deviation)}}  & \multicolumn{3}{c|}{Individual Classifiers}           & \multicolumn{4}{c|}{Decision Level Fusion over Predictions}            \\ \cline{4-10} 
\multicolumn{3}{|l|}{}          &  \shortstack{Softmax \\ Classifier} &  \shortstack{SVM \\ (RBF)} &  \shortstack{SVM \\ (Poly)}   & \shortstack{Fusion \#1 \\ RBF + Poly}   &  \shortstack{Fusion \#2 \\Softmax + RBF}  &  \shortstack{Fusion \#3 \\Softmax + Poly}   & \shortstack{Fusion \#4 \\ All}   \\ \hline
\multicolumn{1}{|c|}{\multirow{21}{*}{\rotatebox[origin=c]{90}{ \shortstack{Individual Performances of \\ Deep Neural Networks  } }}} & \multirow{3}{*}{Inception V3} & DB1 & 87.6 (2.2) &87.6 (2.3) &	87.3 (2.2)  &	87.5 (2) &	87.3 (3.5) &	87.4 (2.3) &	87.7 (2.3)    \\
&   &    DB2 & 83.6 (2.4)&	84.0 (4.4)&	83.6 (4.8) &	83.6 (3.3) &	83.0 (3.4)) &	82.8 (3.3) &	83.7 (3.3)
  \\ 
&   &    DB3 & 84.5 (2.6) &	86.2 (2) &	85.9 (2.6) &	86.4 (2.6) &	86.0 (2) &	86.0 (2.2) &	86.3 (2.2)
  \\ \cline{2-10}

& \multirow{3}{*}{Xception} & DB1 & 86.7 (3.8)&	85.7 (4.2)&	85.6 (3.4)&	85.3 (3.8)&	85.7 (3.8)&	86.4 (3.5)&		86.8 (3.4)
   \\
&     & DB2 & 84.7 (2.1) &	83.7 (2.8) &	83.0 (2.6) &	82.8 (1.5) &	83.9 (1.7) &	83.4 (1.8)&		84.2 (1.2)
  \\ 
&     & DB3 & 85.3 (1.6) &	83.3 (1.8)&	82.1 (3.6)&	82.3 (1.6) &	83.9 (1.6) &	82.9 (0.9) &	84.0 (0.7)
  \\ \cline{2-10}

& \multirow{3}{*}{MobileNet V2} & DB1 & 84.2 (2.8)&	86.8 (2.7)&	86.0 (1.2)&	86.1 (2.1)&	84.7 (1.3)&	84.7 (1.6)&	86.1 (1.1)
    \\
&     & DB2 & 84.5 (2) &	84.7 (2.5) &	84.4 (2.4) &	84.8 (2.6) &	84.6 (2.5) &	84.7 (2.6) &	85.0 (2.5)
  \\ 
&     & DB3 & 85.5 (3.3)&	85.8 (2.8)&	85.6 (2.1) &	85.9 (2.5)&	85.9 (3.2)&	85.9 (2.4) &	86.1 (1.8)
 \\ \cline{2-10}

& \multirow{3}{*}{ResNet50} & DB1 & 86.3 (2.3)&	87.4 (2.5)&	87.0 (2.7)&	87.3 (2.5)&	86.2 (2.6)&	86.4 (2.5)&	87.5 (2.3)
    \\
&     & DB2 & 85.7 (1.2)&	84.9 (5.2)&	82.6 (3.8)&	83.5 (1.8)&	85.0 (1.9)&	83.9 (1.4)&	85.2 (0.8)
  \\ 
&     & DB3 & 85.4 (2.6)&	85.6 (3.1)&	85.2 (3.1)&	85.4 (2.9)&	85.4 (2.6)&	85.4 (2.8)&	85.6 (2.8)
  \\ \cline{2-10}

& \multirow{3}{*}{ResNet101} & DB1 & 85.5 (1.5)&	85.7 (2.1)&	85.5 (1.6)&	85.7 (1.5) &	85.4 (1.8)&	85.4 (2.3)&	85.8 (1.9)
    \\
&     & DB2 & 84.1 (4.4)&	84.8 (3.9)&	83.4 (4.1)&	84.7 (4.3)&	84.8 (3.9)&	84.3 (5)&	84.7 (4)
   \\ 
&     & DB3 & 85.1 (2.8)&	85.3 (3.1)&	84.9 (3)&	85.2 (1.8) &	85.2 (3.1)&	85.1 (2.9)&	85.1 (2.9)
  \\ \cline{2-10}

& \multirow{3}{*}{NasNet} & DB1 & 85.2 (2.5)&	84.8 (2.3)&	84.3 (2.6)&	84.4 (2.3)&	84.8 (2.2)&	84.4 (2.2)&	84.6 (2.1)
    \\
&     & DB2 & 83.3 (2.1) &	81.4 (3.1)&	79.8 (6.4)&	79.7 (3.1)&	81.8 (2.7)&	80.2 (3.2)&	81.8 (3.2)
   \\ 
&     & DB3 & 84.1 (2.6)&	83.6 (2.6)&	82.3 (3)&	82.9 (2.3)&	83.9 (2.6)&	83.4 (2.4)&	83.5 (3.3)
 \\ \cline{2-10}

& \multirow{3}{*}{VGG16} & DB1 & 85.8 (3.1)&	86.3 (2.9)&	85.9 (3.1)&	86.3 (2.3)&	86.0 (3)&	85.9 (2.9)&	86.1 (3.4)
   \\
&     & DB2 & 85.1 (1) &	85.9 (1.1)&	85.2 (1.4)&	86.0 (0.8)&	85.6 (0.6)&	85.4 (1)&	85.9 (1.1)
   \\ 
&     & DB3 & 85.7 (1.8)&	85.9 (2.2)&	85.8 (2.2)&	85.8 (2.2)&	86.0 (2.3)&	86.0 (2.1)&	86.1 (2.1)
\\ \hline
 
\multicolumn{1}{|c|}{\multirow{3}{*}{\rotatebox[origin=c]{0}{ \shortstack{Feature Level \\ Fusion  } }}}                  & \multicolumn{1}{c}{\multirow{3}{*}{\shortstack{Concatenated \\ Vector}}} & DB1 & 90.2 (2.3)&	90.7 (1.7)&	90.3 (2.2)&	90.8 (1.6)&	90.4 (2.1)&	90.4 (2) &	90.8 (1.7)
   \\
&    & DB2 & 87.3 (2.4)&	86.9 (2)&	87.2 (2.2)&		87.1 (2.1)&	87.3 (2)&	87.3 (2.1) &	87.6 (2)
  \\ 
&    & DB3 & 88.7 (3)&	89.2 (2.9)&	89.3 (2.9)&	89.2 (2.8)&	89.0 (2.8)&	89.1 (2.9)&	89.5 (2.7)
  \\  \hline

\end{tabular}}
\label{All_Results_Table}
\end{table}

\begin{figure}[!h]
        \centering
        \includegraphics[width=0.8\linewidth]{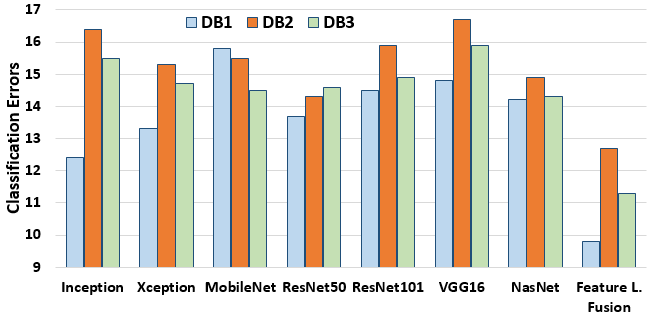}
   \caption{Classification errors of the individual learning and the feature level fusion schemes when the softmax classifier is employed.}
    \label{Network_CNN_Errors}
\end{figure}

Regarding the effect of feature level fusion, the bottom row group of Table \ref{All_Results_Table} and the Figure \ref{Network_CNN_Errors}, in which the error values obtained from the three COVID-19 databases for the individual softmax classifier based learning models plus concatenated feature vector can be investigated. As seen in Figure \ref{Network_CNN_Errors}, the error values, which were obtained from the deep feature vector formed by using feature level fusion, are significantly lower than individual softmax classifier performance by reaching 9.8\%, 12.7\% and 11.3\% errors for the DB1, DB2 and DB3 respectively. As understood from Table \ref{All_Results_Table} and the Figure \ref{Network_CNN_Errors}, not a specific individual deep feature set (extracted by using a specific CNN) has outperformed the others for all three databases. This situation indicates that there is a significant need for ensemble learning which may pave the way for the complementary information achievement. It should also be noted that the error values for DB1 and DB3 were even further reduced by 0.5\% and 0.6\% respectively when RBF and polynomial kernel based multistage learning algorithms were applied.   

\begin{figure*}[!t]
    \centering
        \centering
        \includegraphics[width=6.1in]{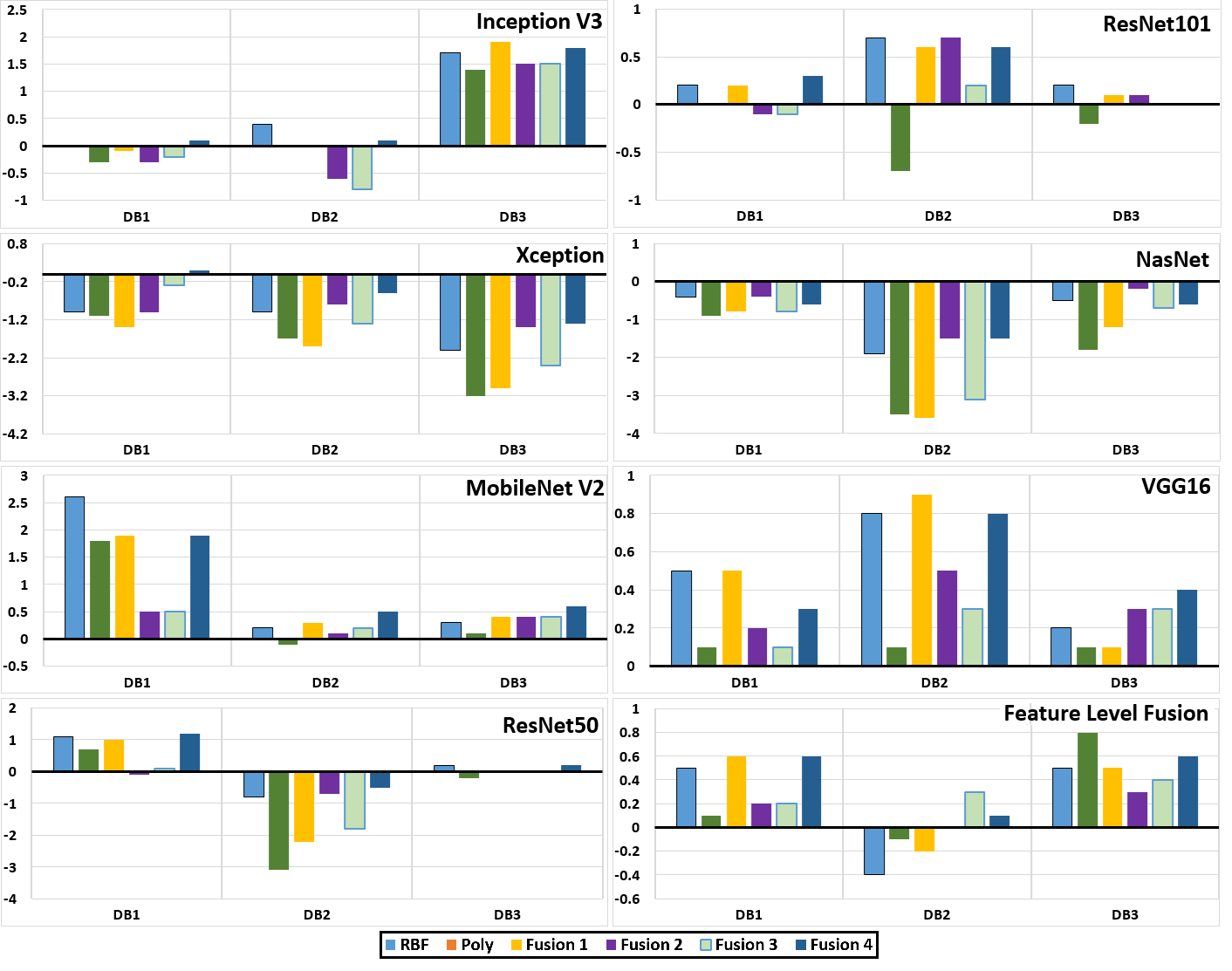}
        \caption{The accuracy variations compared to CNNs when multistage learning and/or majority voting is applied are presented.}
        \label{DecisionLevelFusion}
\end{figure*}

The contribution of decision level fusion can be investigated by using the right-side of Table \ref{All_Results_Table} and the Figure \ref{DecisionLevelFusion}.
In Figure \ref{DecisionLevelFusion}, the conventional classification performance of the softmax classifier (as it is used in traditional CNN based learning) was chosen as the reference baseline performance for seven CNN based deep feature extraction schemes. For comparison, the increments or decrements seen in the accuracy values obtained by the multistage SVM based learning and the decision level fusion were represented for each deep feature set plus the concatenated feature vector (obtained by the feature level fusion). When the Table \ref{All_Results_Table} is investigated, it is seen that the highest accuracy values within the entire test set combinations were obtained as 90.8\%, 87.6\% and 89.5\% for the DB1, DB2 and DB3 respectively, when the fourth decision level fusion approach, including the majority voting of hard labels obtained by softmax classifier, RBF and polynomial SVMs, was employed. As illustrated in Figure \ref{DecisionLevelFusion}, almost for all multistage SVM based learning and decision level fusion cases applied to MobileNetV2 and VGG16 based deep features, up to 2.5\% increase in accuracy rate was achieved. On the contrary, approximately all the accuracy values obtained by decision level fusion, when they were applied to Xception and NasNet based deep features, were lower than the baseline softmax classifier performance. In accordance with the remaining ResNet50, ResNet101 and  InceptionV3 based deep features, neither the positive nor the negative effect of multistage learning and decision level fusion was clearly seen. For instance, up to 2\% increase in the accuracy values  was seen for the InceptionV3 based scenarios in DB3, while slight improvements have been achieved by using ResNet101 based scenarios for DB2.     

\begin{figure*}[!t]
    \centering
      \includegraphics[width=6.1in]{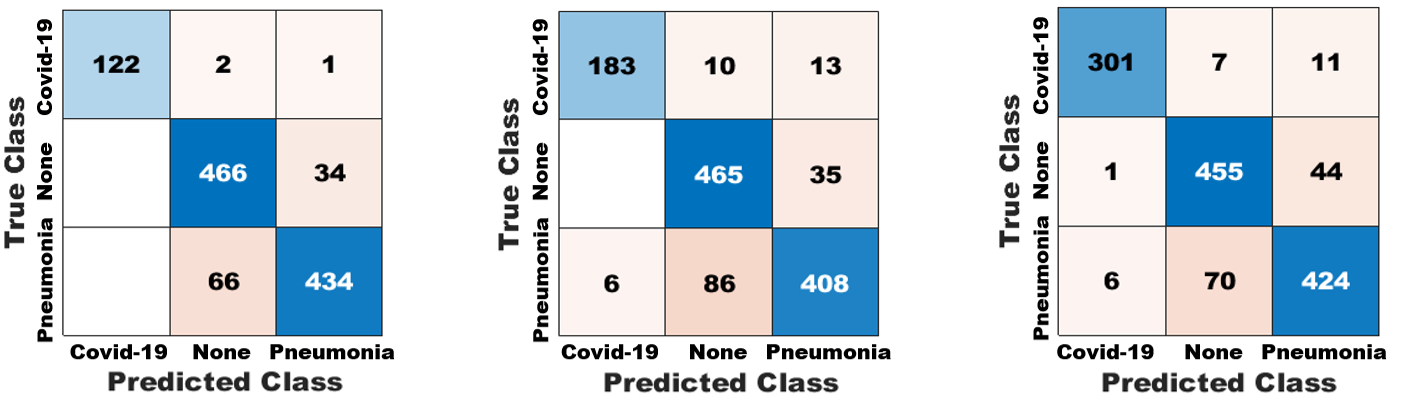}    \caption{Confusion Matrices obtained from Fusion \#4 strategy (Left DB1, Middle DB2, Right DB3)}
    \label{ConfMats}
\end{figure*}

As alternative objective evaluation criteria, the confusion matrix based metrics were calculated to be able to show the performance of proposed approach. For doing this, the true positive (TP), true negative (TN), false positive (FP), and false negative (FN) values were obtained for each database. The confusion matrices obtained by the Fusion \#4 strategy applied to 3 databases were given as Figure \ref{ConfMats} for further understanding. After the confusion matrices were obtained, 3 objective evaluation metrics were calculated as follows:

\begin{equation}
Precision = \frac{TP}{TP+FP}
\end{equation}
\begin{equation}
Recall = \frac{TP}{TP+FN}
\end{equation}
\begin{equation}
F1\, score = 2\: \frac{Precision \times Recall}{Precision + Recall}
\end{equation}

\begin{figure*}[!t]
    \centering
    \begin{subfigure}[b]{1\textwidth}
        \centering
        \includegraphics[width=6in]{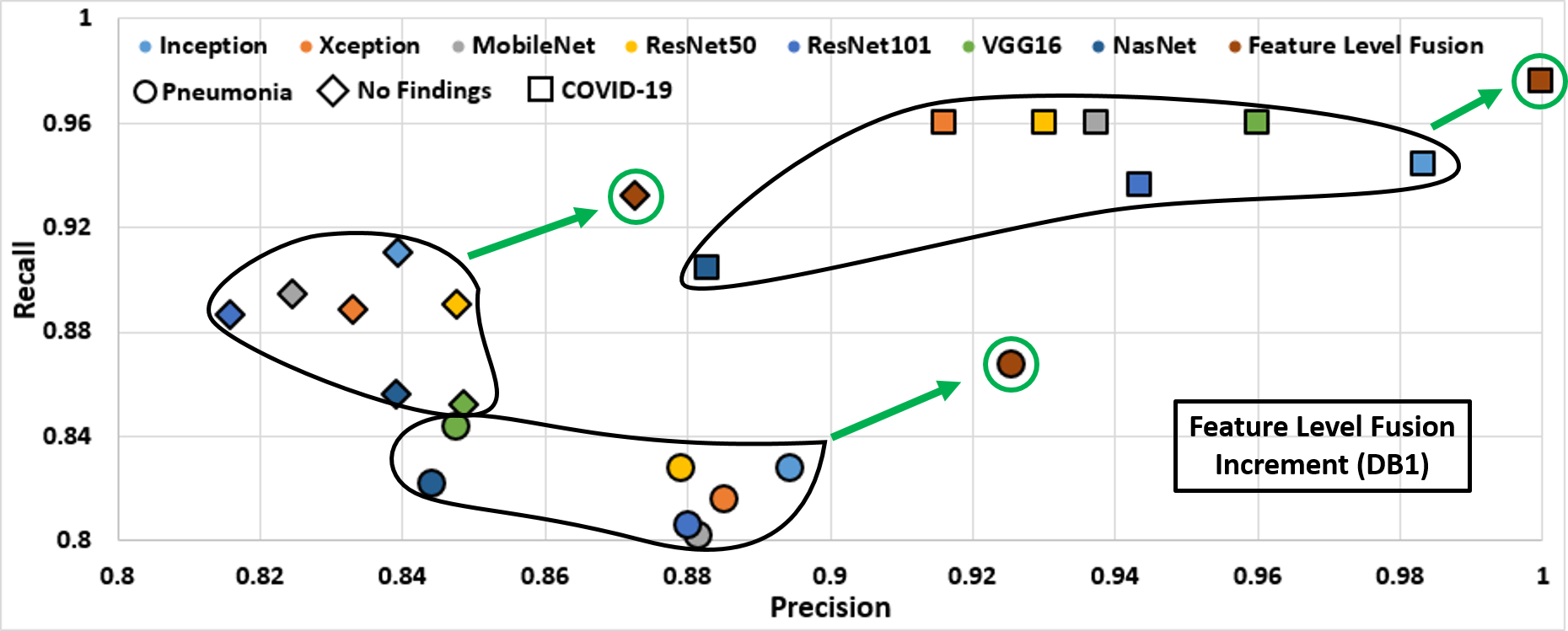}
        \caption{}
    \end{subfigure}%
  
    \begin{subfigure}[b]{1\textwidth}
        \centering
        \includegraphics[width=6in]{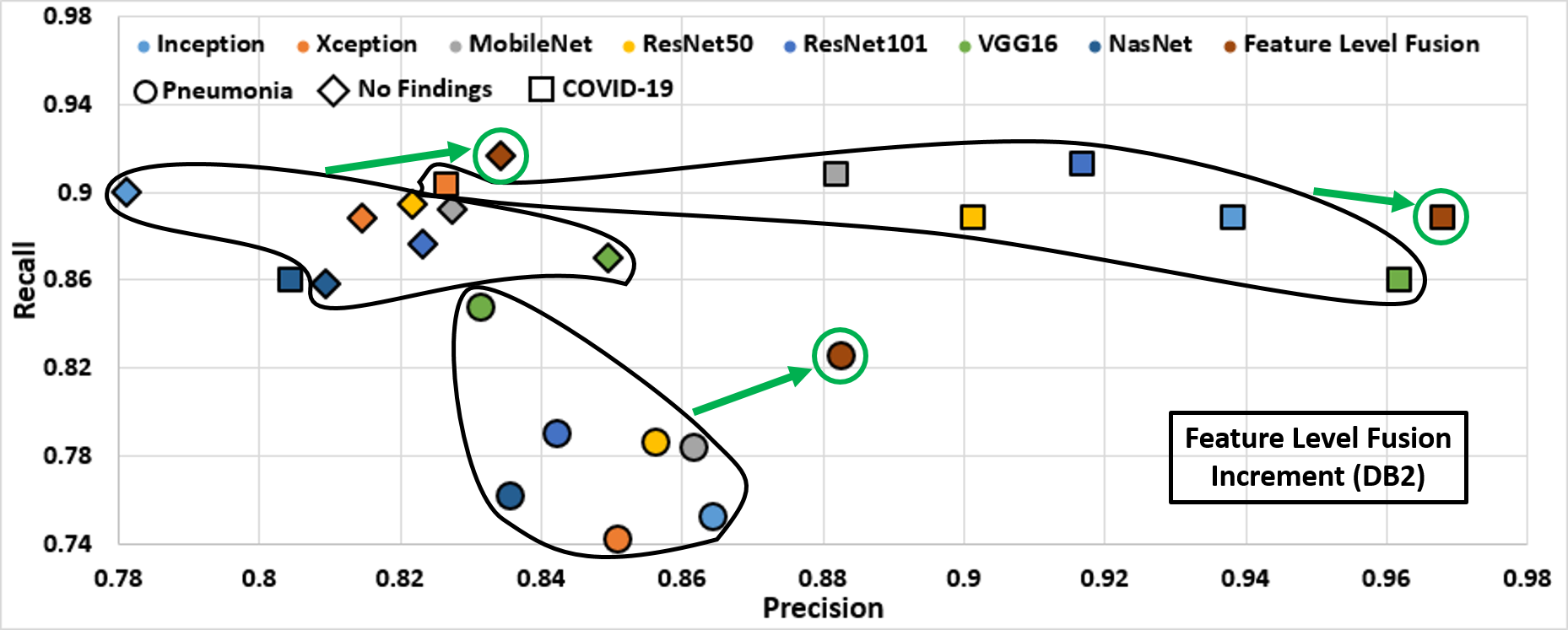}
        \caption{}
    \end{subfigure}%

    \begin{subfigure}[b]{1\textwidth}
        \centering
        \includegraphics[width=6in]{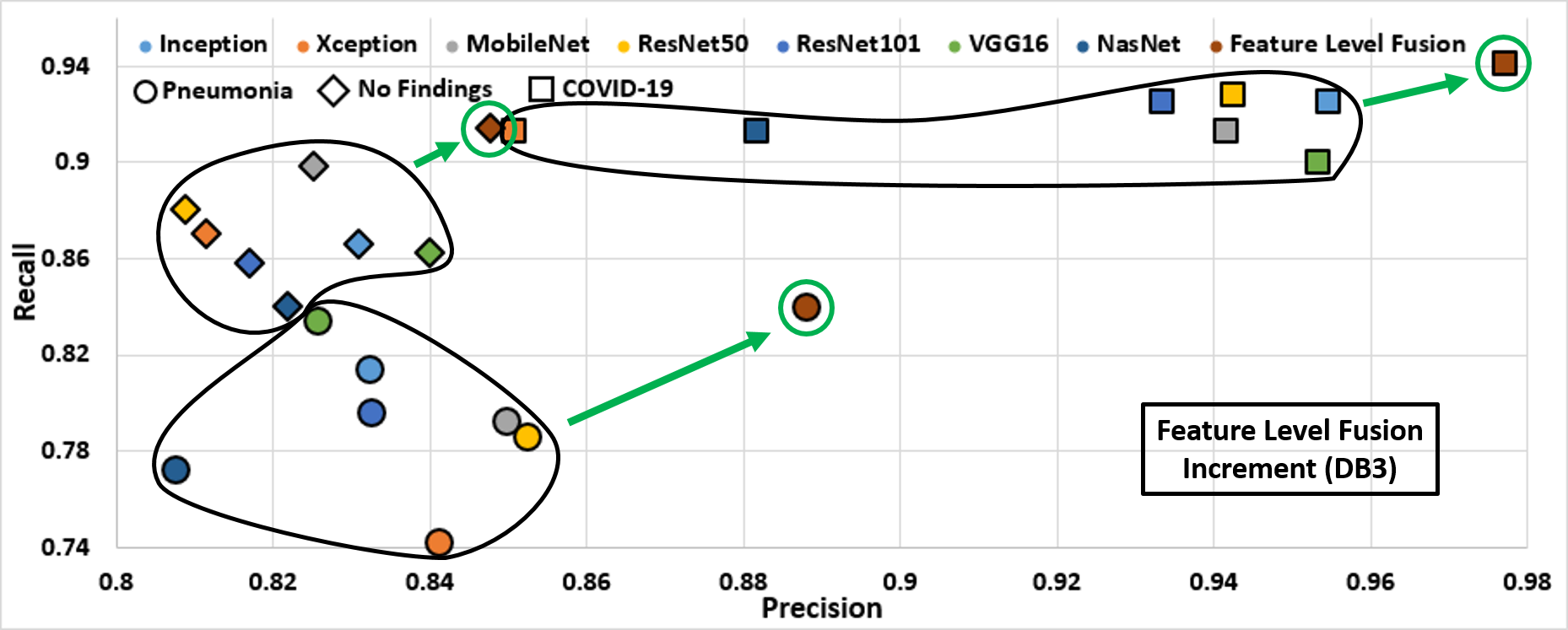}
        \caption{}
    \end{subfigure}%
    \caption{Obtained precision and recall values of Fusion \#4 strategy for each individual CNN and concatenated feature vector.}
    \label{PrecisionVSRecall}
\end{figure*}

Among these, the precision emphasizes how precise the learning model is out of those predicted positive samples, how much of the predicted positives are actual positive. The precision is an important parameter to determine when the costs of FP predictions is high. Moreover, the recall measures how much of the actual positive samples are captured by the model by labeling it as positive (TP). The recall is an essential parameter when there is a high cost associated with FN samples. The behaviour of precision vs recall of the COVID-19, pneumonia and no-finding classes obtained by using majority voted decisions, described as Fusion \#4, of individual deep feature sets (obtained by a specific CNN) plus the concatenated feature vector (obtained by the feature level fusion) is given in Figure \ref{PrecisionVSRecall}. It is seen that the precision and recall values obtained by the concatenated feature vector were higher than the individual deep feature sets in almost all cases. As presented in Table \ref{All_Results_Table}, the highest accuracy values were obtained when the Fusion \#4 strategy was applied for all three databases. The obtained precision and recall values for Fusion \#4 strategy is also depicted in Table \ref{ConfMats_based_metrics} to go in deeper investigation. As seen in this table, in almost all classes and databases, the highest precision, recall and F1-scores were obtained for the COVID-19 class which has the highest priority in our classification problem. In addition, an important evaluation metric named as Kappa, which is a statistical measure of inter-annotator agreement for categorical items by comparing an observed accuracy with an expected accuracy \cite{mchugh2012interrater}, was given for all databases in Table \ref{ConfMats_based_metrics}. As mentioned in \cite{landis1977application}, the Kappa values greater than 0.80 are called almost perfect classification. Hence, the obtained Kappa values ($0.845$, $0.798$ and $0.835$ for DB1, DB2 and DB3 respectively) shows the success of proposed approach following Fusion \#4 strategy in COVID-19 diagnosis problem. As a final point to remark, all the F1-measure values indicating how precise the classifier is (what percentage of the samples assigned to a certain class is classified correctly), as well as how robust it is (what percentage of the samples belonging to a certain class is classified correctly), were quite high for the COVID-19 class, showing the success of proposed Fusion \#4 strategy.

\begin{table*}[!h]
\caption{The detailed presentation of evaluation metrics for the Fusion \#4 strategy.}
\scriptsize
\centering
\resizebox{\textwidth}{!}{
\begin{tabular}{clccccccccc}
     &     &  \multicolumn{3}{c}{DB1 (\cite{ozturk2020automated})}      & \multicolumn{3}{c}{DB2 (\cite{ozturk2020automated} + \cite{cohen2020covid})} & \multicolumn{3}{c}{DB3 (\cite{ozturk2020automated} + \cite{cohen2020covid} + \cite{Agchung})} \\ \cmidrule(lr){3-5} \cmidrule(lr){6-8}  \cmidrule(lr){9-11}
          & & COVID-19 & No Findings & Pneumonia & COVID-19 & No Findings & Pneumonia & COVID-19 & No Findings & Pneumonia   \\  \cmidrule(lr){3-5} \cmidrule(lr){6-8}  \cmidrule(lr){9-11}
\multicolumn{1}{c}{\multirow{5}{*}{\rotatebox[origin=c]{90}{ \shortstack{Feature Level \\ Fusion} }}} & Precision &  100 &  87.26  &  92.53  &  96.82  &  82.88  &   89.47  & 97.72 & 85.52 & 88.51   \\
& Recall    & 97.6  &  93.2  & 86.8 &  88.83     &  93  & 81.6 & 94.35 & 91 & 84.8 \\
& F-Score   &  98.78 &  90.13  &  89.57 & 92.65 & 87.65 & 85.35 & 96.01 & 88.17 & 86.62  \\  \cmidrule(lr){3-5} \cmidrule(lr){6-8}  \cmidrule(lr){9-11}
& Accuracy  & \multicolumn{3}{c}{90.84}                & \multicolumn{3}{c}{87.56}   & \multicolumn{3}{c}{89.46}        \\
& Kappa     & \multicolumn{3}{c}{0.845}     & \multicolumn{3}{c}{0.801} & \multicolumn{3}{c}{0.839}            \\  \cmidrule(lr){3-11} 
\end{tabular}}
\label{ConfMats_based_metrics}
\end{table*}

\section{Discussion and Conclusion}
\label{DiscussionandConclusion}

Although the RT-PCR is the most common technique to diagnose COVID-19, chest radiography based approaches have been extensively used as complementary diagnosis tools due to the low-sensitivity drawback of RT-PCR especially seen in the early stage of COVID-19. The X-ray scanning has been preferred as the primary radiography based imaging approach in COVID-19 detection due to its fast imaging speed, low cost and low dosage exposing of radiation compared to CT. However, the interpretation success of X-ray images strongly depends on the radiologist's experience and visual inspection of the X-ray images belonging to several patients takes significant time and effort. In order to increase the objectivity of the X-ray imaging interpretation and decrease the required time and effort, CAD systems have been used as supporting decision mechanisms in the detection of COVID-19 cases. In this respect, several studies employing deep networks as the decision tool were published lately as depicted in Table \ref{Literature_Comparision_Table}. The performance of our proposed study is compared with the previous studies in terms of the; i) the number of COVID-19 image samples existing in the employed database, ii) the number of classes that are tried to be separated, iii) the accuracy value plus the precision and recall metrics. As seen in Table \ref{Literature_Comparision_Table}, the number of employed X-ray samples for model generation is very low in \cite{sethy2020detection}, \cite{narin2020automatic} and \cite{hemdan2020covidx}, which have used only 50, 100 and 100 X-ray images respectively. In such systems, which have few number of samples for training and testing, the learning model usually memorizes the data resulting in over-fitting. On the other hand, although nearly sufficient number of X-ray samples exist in \cite{mahmud2020covxnet}, \cite{oh2020deep}, \cite{wang2020covid}, \cite{apostolopoulos2020covid}, \cite{brunese2020explainable} and \cite{zhang2020covid}, the ratio of COVID-19 samples are very low compared to the distribution of the remaining classes. However, most of the learning models tend to work on balanced class distributions or equal misclassification costs, and the performance of these learning methods can be significantly compromised when imbalanced data sets, like the employed COVID-19 vs non COVID-19 distributions seen in \cite{mahmud2020covxnet,oh2020deep,wang2020covid,apostolopoulos2020covid,brunese2020explainable,zhang2020covid}, are used. Therefore, in our study, the employed data bases were progressively created, starting from the usage of samples given in \cite{ozturk2020automated} as DB1, till minimum imbalance between employed classes was achieved in DB3. As seen in Table \ref{Literature_Comparision_Table}, our method has outperformed \cite{ozturk2020automated}, \cite{mahmud2020covxnet} and \cite{zhang2020covid} in terms of accuracy, precision and recall metrics, while our algorithm provides competitive performance compared to \cite{oh2020deep}, \cite{wang2020covid}, \cite{apostolopoulos2020covid} and \cite{brunese2020explainable} in terms of accuracy. It should be noted that our approach applied to DB1 is having the same number of image samples and same cross-validation strategy compared to \cite{ozturk2020automated}, while a similar 5-fold cross-validation with different number of X-ray samples was carried out in \cite{mahmud2020covxnet} and \cite{zhang2020covid}. Furthermore, the precision values obtained by using our method were significantly higher than \cite{ozturk2020automated}, \cite{mahmud2020covxnet}, \cite{oh2020deep} and \cite{zhang2020covid}, while higher performance was achieved in terms of recall compared to \cite{wang2020covid}, \cite{apostolopoulos2020covid} and \cite{brunese2020explainable}.   

\begin{table*}[t]
\centering
\small
\caption{Performance comparison of related works on COVID-19 detection problem with the proposed method.}
\label{litCompare}
\resizebox{\textwidth}{!}{
\begin{tabular}{lcclccc}
 Paper & Architecture   &\begin{tabular}[c]{@{}c@{}}Total Number \\ of Images \end{tabular}  & \begin{tabular}[c]{@{}c@{}}Class Names and \\ number of Images\end{tabular}    &  Accuracy (\%) & \begin{tabular}[c]{@{}l@{}} Precision (\%) \\ (COVID-19) \end{tabular} &  \begin{tabular}[c]{@{}l@{}} Recall (\%) \\ (COVID-19) \end{tabular} \\ \hline \hline

\multirow{3}{*}{ \begin{tabular}[l]{@{}l@{}} Ozturk et al.\\ 2020 \cite{ozturk2020automated} \end{tabular}  }   & \multirow{3}{*}{ \begin{tabular}[c]{@{}c@{}} DarkCovidNet \end{tabular}  }    & \begin{tabular}[c]{@{}c@{}} 1125 \end{tabular}      & \begin{tabular}[l]{@{}l@{}} COVID-19 (125) \\ No Findings (500) \\ Pneumonia (500) \end{tabular}    &  87.02 & 80.7  &   97.87 \\ \cline{3-7}

&  & \begin{tabular}[c]{@{}c@{}} 625 \end{tabular}      & \begin{tabular}[l]{@{}l@{}} COVID-19 (125) \\ No Findings (500)  \end{tabular}    &  98.08 & 97.97 & 90.65   \\ \hline

\multirow{1}{*}{ \begin{tabular}[l]{@{}l@{}} Mahmud et al.\\ 2020 \cite{mahmud2020covxnet} \end{tabular}  }   & \multirow{1}{*}{ \begin{tabular}[c]{@{}c@{}}  CovXNet \end{tabular}  }    & \multirow{1}{*}{ \begin{tabular}[c]{@{}c@{}} 5856 \end{tabular}  }    & \begin{tabular}[l]{@{}l@{}} COVID-19 (305) \\ No Findings (1583) \\ Viral Pneumonia (1493) \\ Bacterial Pneumonia (2780) \end{tabular}    &  90.2 & 90.8  &   89.9 \\ \hline

\begin{tabular}[l]{@{}l@{}} Oh et al.\\ 2020 \cite{oh2020deep} \end{tabular}    & \multirow{1}{*}{ \begin{tabular}[c]{@{}c@{}}  A patch-based
ResNet18 \end{tabular}  }    & \multirow{1}{*}{ \begin{tabular}[c]{@{}c@{}} 15043 \end{tabular}  }    & \begin{tabular}[l]{@{}l@{}} COVID-19 (180) \\ Pneumonia (6012) \\ Normal (8851)  \end{tabular}    &  91.9 &  76.9 & 100 \\ \hline

\multirow{3}{*}{ \begin{tabular}[l]{@{}l@{}} Waheed et al.\\ 2020 \cite{waheed2020covidgan} \end{tabular}  }   & \multirow{3}{*}{ \begin{tabular}[c]{@{}c@{}} Augmentation by CovidGAN \\ Classification by VGG16 \end{tabular}  }    & \begin{tabular}[c]{@{}c@{}} 1124 \\ No Augmentation \end{tabular}      & \begin{tabular}[l]{@{}l@{}} COVID-19 (403) \\ Normal (721) \end{tabular}    &  85 & 89  &   69 \\ \cline{3-7}

&  & \begin{tabular}[c]{@{}c@{}} 3260 \\ With Augmentation \end{tabular}      & \begin{tabular}[l]{@{}l@{}} COVID-19 (1741) \\ Normal (1519)  \end{tabular}    &  95 & 96 & 90   \\ \hline

 \begin{tabular}[l]{@{}l@{}} Wang et al.\\ 2020 \cite{wang2020covid} \end{tabular}    & \multirow{1}{*}{ \begin{tabular}[c]{@{}c@{}}  COVID-Net \end{tabular}  }    & \multirow{1}{*}{ \begin{tabular}[c]{@{}c@{}} 13975 \end{tabular}  }    & \begin{tabular}[l]{@{}l@{}} COVID-19 (358) \\ Pneumonia (5551) \\ Normal (8066)  \end{tabular}    &  93.3 &  98.91 & 91 \\ \hline

\begin{tabular}[l]{@{}l@{}} Ioannis and Mpeslana\\ 2020 \cite{apostolopoulos2020covid} \end{tabular}     & \multirow{1}{*}{ \begin{tabular}[c]{@{}c@{}}  VGG-19 \end{tabular}  }    & \multirow{1}{*}{ \begin{tabular}[c]{@{}c@{}} 1428 \end{tabular}  }    & \begin{tabular}[l]{@{}l@{}} COVID-19 (224) \\ Normal (504) \\ Pneumonia (700) \end{tabular}    &  93.48 &  98.75 &  92.85  \\ \hline

\begin{tabular}[l]{@{}l@{}} Sethy and Behera \\ 2020 \cite{sethy2020detection} \end{tabular}     & \multirow{1}{*}{ \begin{tabular}[c]{@{}c@{}}  ResNet-50 + SVM \end{tabular}  }    & \multirow{1}{*}{ \begin{tabular}[c]{@{}c@{}} 50 \end{tabular}  }    & \begin{tabular}[l]{@{}l@{}} COVID-19 (25) \\ Non-Covid (25)  \end{tabular}    &  95.38 & 93.47  &   97.24 \\ \hline

\begin{tabular}[l]{@{}l@{}} Narin et al.\\ 2020 \cite{narin2020automatic} \end{tabular}     & \multirow{1}{*}{ \begin{tabular}[c]{@{}c@{}}  ResNet-50  \end{tabular}  }    & \multirow{1}{*}{ \begin{tabular}[c]{@{}c@{}} 100 \end{tabular}  }    & \begin{tabular}[l]{@{}l@{}} COVID-19 (50) \\ Non-Covid (50)  \end{tabular}    &  98 & 100 & 96   \\ \hline

\begin{tabular}[l]{@{}l@{}} Hemdan et al.\\ 2020 \cite{hemdan2020covidx} \end{tabular}     & \multirow{1}{*}{ \begin{tabular}[c]{@{}c@{}}  COVIDX-Net  \end{tabular}  }    & \multirow{1}{*}{ \begin{tabular}[c]{@{}c@{}} 50 \end{tabular}  }    & \begin{tabular}[l]{@{}l@{}} COVID-19 (25) \\ Non-Covid (25)  \end{tabular}    &  90 & 83  & 100   \\ \hline


\begin{tabular}[l]{@{}l@{}} Brunese et al. \\ 2020 \cite{brunese2020explainable} \end{tabular}     &  \begin{tabular}[c]{@{}c@{}}  Cascade Analyse \\ by Multiple CNN models \\ (Only Second Model Results) \end{tabular}      & \begin{tabular}[c]{@{}c@{}} 3003 \end{tabular}     & \begin{tabular}[l]{@{}l@{}} COVID-19 (250) \\ Other(2753)  \end{tabular}    &  97 &  94 &   87 \\ \hline

\begin{tabular}[l]{@{}l@{}} Zhang et al. \\ 2020 \cite{zhang2020covid} \end{tabular}     &  \begin{tabular}[c]{@{}c@{}}  Confidence-aware \\ Anomaly
Detection (CAAD) \end{tabular}      & \begin{tabular}[c]{@{}c@{}} 43583 \end{tabular}     & \begin{tabular}[l]{@{}l@{}} COVID-19 (106) \\ Normal(107) \\ Viral Pneumonia (5977) \\ Healthy + Non-Viral (37393)  \end{tabular}    &  72.77 &  71.7 &  73.83 \\ \hline

\multirow{6}{*}{\begin{tabular}[l]{@{}l@{}} Proposed \\ Study \end{tabular}  }   & \multirow{1}{*}{ \multirow{6}{*}{ \begin{tabular}[c]{@{}c@{}}  Deep Feature and \\ Decision Level \\ Fusion ()  \end{tabular}  }   } & \multirow{1}{*}{ \begin{tabular}[c]{@{}c@{}} 1125 \end{tabular}  }    & \begin{tabular}[l]{@{}l@{}} COVID-19 (125) \\ No Findings (500) \\ Pneumonia (500) \end{tabular}    &  90.84 & 100  &  97.6  \\ \cline{3-7}

&   & \multirow{1}{*}{ \begin{tabular}[c]{@{}c@{}} 1206 \end{tabular}  }    & \begin{tabular}[l]{@{}l@{}} COVID-19 (206) \\ No Findings (500) \\ Pneumonia (500) \end{tabular}    &  87.56 & 96.82  & 88.83   \\ \cline{3-7}

&   & \multirow{1}{*}{ \begin{tabular}[c]{@{}c@{}} 1319 \end{tabular}  }    & \begin{tabular}[l]{@{}l@{}} COVID-19 (319) \\ No Findings (500) \\ Pneumonia (500) \end{tabular}    &  89.46 &  97.72 &  94.35  \\ 

\hline \hline  
\end{tabular}}
\label{Literature_Comparision_Table}
\end{table*}

As seen in Table \ref{All_Results_Table} and the Figure \ref{Network_CNN_Errors}, none of the individual learning models has been significantly outperformed the others (The statistical significance analysis results of the tested approaches are given in the Supplementary Materials as Table S1 and S2). However, accuracy improvements up to 3\% were achieved when feature level fusion has been applied to obtained deep features. When the multistage learning and decision level fusion approaches are investigated, it is seen that the accuracy rises up to 2.5\% and 1\% have been obtained for the deep features extracted by using MobileNetV2 and VGG16 respectively. The supportive effect of SVM usage and majority voting for these two CNNs can be related to their sizes, which are the cause of possible underfitting and overfitting. As mentioned in \cite{zhang2018shufflenet}, small networks such as the MobileNetV2 usually suffer from underfitting, while very large models such as the VGG16 may have trouble with overfitting \cite{howard2017mobilenets}. However, a learner such as SVM, which is good at producing optimal decision surfaces even there is noise on the data, can have positive effect on the classification accuracy similar to our case. On the contrary, same multistage learning and majority voting strategy did not work well, resulting accuracy reductions for the deep featured obtained by Xception and NasNet. When the architecture of NasNet is investigated, it is seen that the NasNet was constructed by a neural architecture search based optimization carried out by using reinforcement learning. As a result of this process, the well-designed scalable and convolutional cells are defined in the optimum way, resulting in an architecture that is prone to produce robust features as in our case. In a similar way, in the Xception, the usage of depthwise separable convolutions paves the way of efficiently usage of model parameters producing stronger features. Hereby, the cascade connection of the SVMs to the last FC layer of Xception and NasNet plus the usage of majority voting has no supportive effect in classification. So, the network based discrimination is more than enough for these two CNNs. 

Another important fact that needs to be discussed about our proposed system, in which the Fusion \#4 strategy was applied, is the obtained high precision and recall values. The precision value is directly related with the number of FP samples and low precision in COVID-19 means high number of healthy subjects that are misdiagnosed as COVID-19. An early quarantine measure applied to COVID-19 patients is employed as the fundamental disease control strategy across the countries \cite{rubin2020psychological}. Apart from the physical damages, the quarantine may cause dramatic psychological effects on the mental health. In previous studies, it was reported that the psychological impact of quarantine can vary from immediate effects such as irritability, fear of spreading infection to family members, confusion, anger, loneliness, anxiety, frustration, denial, insomnia, despair, depression, to extremes of consequences including suicide \cite{dubey2020psychosocial,brooks2020psychological,barbisch2015there}. Therefore, the FP samples frequently seen in a COVID-19 detection system may cause significant undesired psychological and social consequences. However, as seen in Table \ref{ConfMats_based_metrics}, the proposed system has precision values, belonging to COVID-19 class, as 100\%, 96.82\% and 97.72\% for the DB1, DB2 and DB3 respectively showing its almost perfect FP sample reduction performance. The recall metric, which is directly connected to FN samples, is also essential in COVID-19 detection because of the high cost associated with FN samples. Misdiagnosing a COVID-19 patient may cause dramatic consequences due to the very easy and fast transmission  mechanism  of  the  SARS-CoV2. The subject misdiagnosed as normal can spread the disease to his/her close environment in a very short time resulting in new patients who are ready to spread the disease further. However, thanks to our proposed approach, high recall values, reaching up to 97.6\% and 94.35\% in DB1 and DB3 respectively, were obtained by using the Fusion \#4 strategy.

\begin{figure*}
    \centering
    \includegraphics[width=\textwidth]{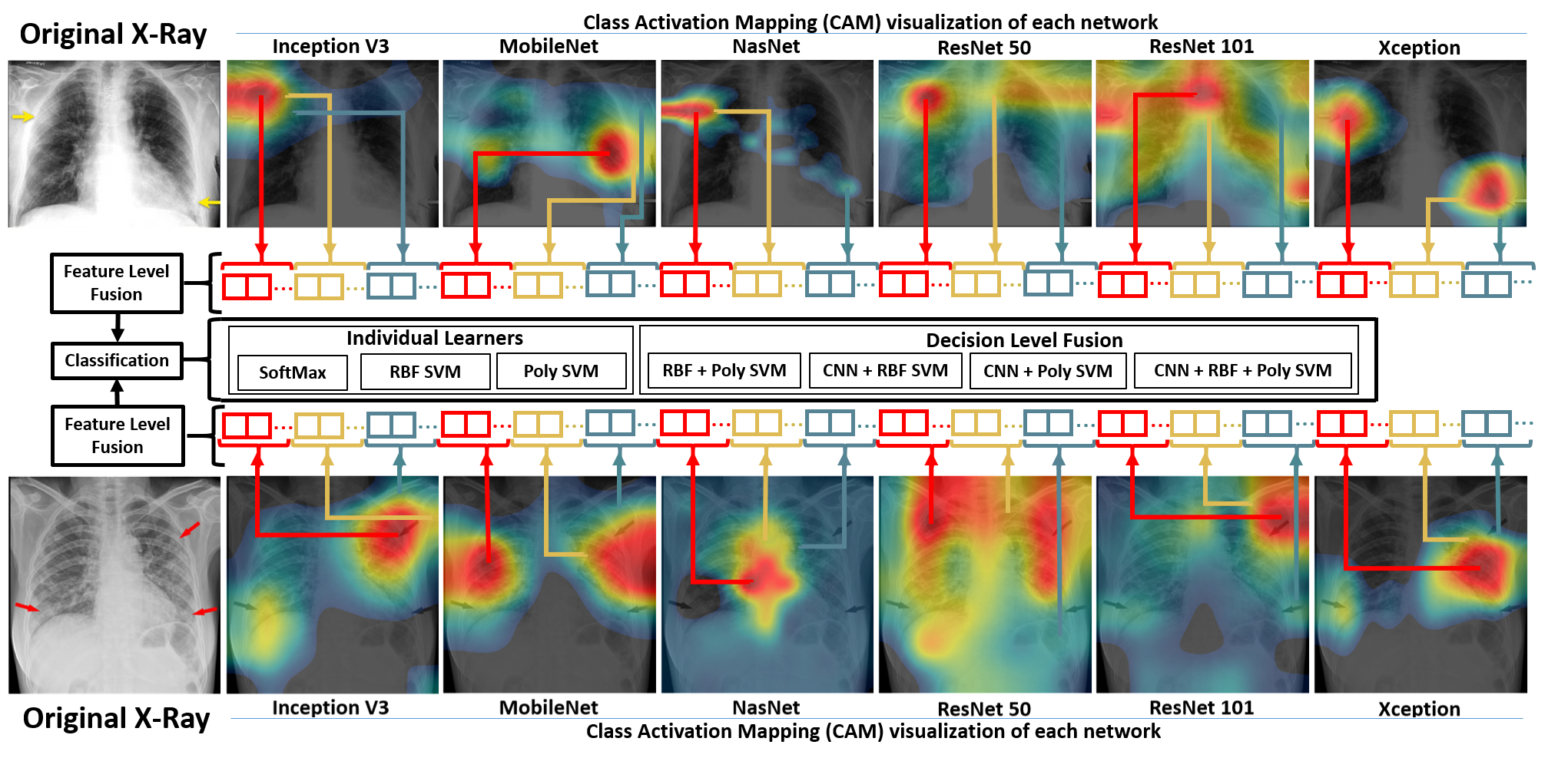}
    \caption{CAM visualizations of two patients obtained by six CNNs (top and bottom rows) and the flow-chart of employed feature level fusion (middle row).}
    \label{fig:CAM_figure}
\end{figure*}

\begin{figure*}[h]
    \centering
    \includegraphics[width=\textwidth]{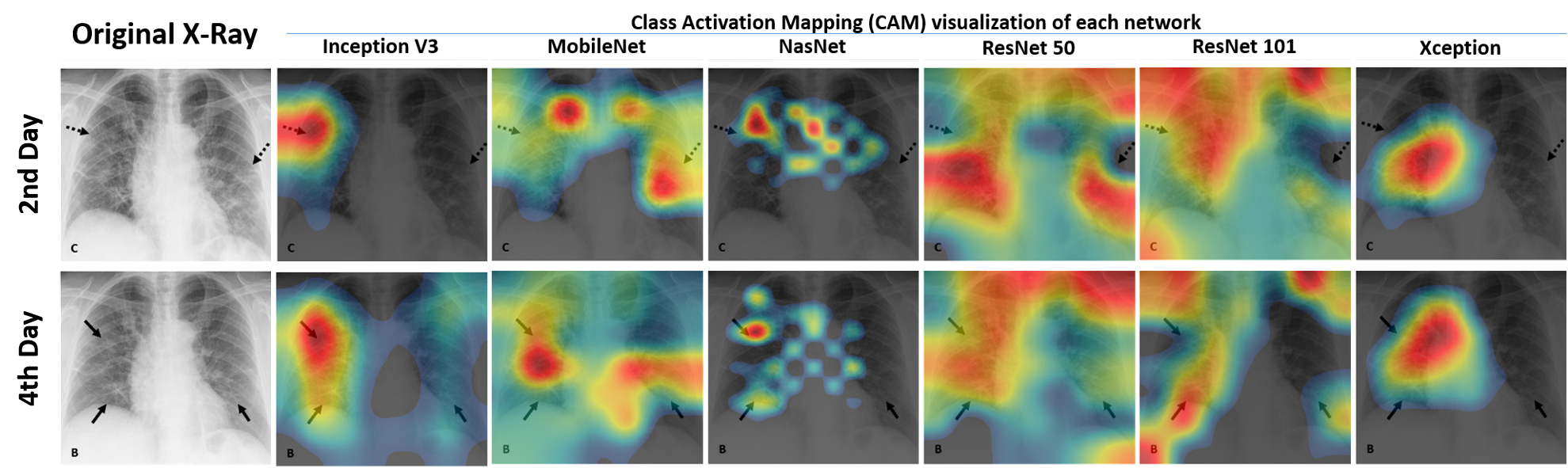}
    \caption{CAM visualizations of the same patient on the second and fourth day of diagnosis.}
    \label{fig:CAM_figure_Consistency}
\end{figure*}

Although the deep learning approaches have enabled unprecedented breakthroughs in medical image analysis, the interpretable modules are sacrificed for uninterpretable ones that achieve higher performance through greater abstraction (more layers) and tighter integration (end-to-end training) in CNNs \cite{selvaraju2017grad}. However, in \cite{zhou2016learning}, the Class Activation Mapping (CAM) technique, which is a way of producing visual explanations of the predictions of deep learning models \cite{arun2020assessing}, was proposed to make the CNNs more transparent and explainable. By using the CAM technique, useful knowledge about the employed prediction regions in the COVID-19 detection problem can be investigated. For example, the failure regions can be visually identified for the wrongly classified samples and necessary modifications in the learning models can be applied towards the most fruitful research directions. Besides, for a deep model, which is very strong in diagnosis, the CAM technique can visually identify the lung consolidation patterns as a supportive diagnostic tool for  doctors. In Figure \ref{fig:CAM_figure}, two CAMs obtained from COVID-19 samples are given with the aim of visual validation of employed CNNs. In the CAMs, the red color highlights the lung regions where the employed CNN model focuses on (activating around those patterns) most during the discrimination. In Figure \ref{fig:CAM_figure}, upper row, the CAMs obtained with six CNNs, excluding VGG16 due to inability of representing its CAM by using employed approach, for a 83 year old male having mitral insufficiency, pulmonary hypertension and atrial fibrillation with COVID-19 infection, can be seen. In this patient, Ground-glass opacification (GGO) and consolidation in the right upper lobe and left lower lobe is seen as the indicators of COVID-19. The InceptionV3 and ResNet50 have correctly localized the right upper lobe pattern, while missing the left lower lobe. However, the Xception has successfully detected both two pathological regions with high spatial resolution. In the bottom row, the CAMs of a 53-year-old female, whose X-ray contains multifocal patchy opacities in both lungs, was depicted. This case is a good example to see the effect of feature level fusion of different CNNs because of the existing three separate opacity patterns. While the MobileNet has strong focus on right side single pattern, the InceptionV3, ResNet50 and Xception has low activation on right side. However, the ResNet101 and InceptionV3 have highly focused on left side upper pattern, while the Xception and MobileNet has significant activity near the left side lower pattern. When the complementary effect of these CAMs is considered, it is obvious that the fusion of features obtained by these CNNs would have higher discriminating power. In the middle part of Figure \ref{fig:CAM_figure}, a flowchart explaining how the features obtained from various CNNs are concatenated is given for further understanding. Additionally, in Figure \ref{fig:CAM_figure_Consistency}, X-ray images belonging to the same patient with bilateral GGO are shown. The image in upper row is taken on the second day of diagnosis while the bottom row X-ray image is taken on the fourth day. As it can be seen, the active regions belonging to a specific CNN are consistent and not dramatically changing towards second and fourth day images.

In future research, we aim to focus on following research paths related with COVID-19 for further improvement; i) a different version of the feature level fusion, in which the features obtained from the various layers of the same CNN are concatenated, can be employed instead of the fusion of features obtained from the last FC layer of different type CNNs. By doing that diverse features, which contain more semantic information in the top layers and more low-level information in bottom layers, can be combined to provide more discriminative information. ii) since the outbreak is recent, the number of COVID-19 X-ray images, which can be used in CAD system design studies, is very limited. Even though there exists a recent study \cite{waheed2020covidgan} that uses Generative Adversarial Network (GAN) for increasing the number of training samples, the performance can be improved by using Progressive Growing GAN \cite{karras2017progressive} for augmentation. Besides, the quality of artificial COVID-19 samples can be improved by integrating more labeled data into the learning process by using GANs. iii) the Canonical Correlation Analysis (CCA) \cite{10.2307/2333955,kettenring1971canonical}, which aims at measuring linear relationships between two sets of variables by using the within-set and between-set sample covariance matrices, can be employed as a feature fusion approach instead of simple concatenation of deep features. By utilizing the multi-view features (the deep features extracted from different CNNs and/or from the different layers of the same CNN), more discriminating features having maximized correlation between various sets can be attained with the hope of performance increase in COVID-19 detection. iv) the hyperparameters, which are adjusted prior to the learning process and affect how the learning algorithm fits the model to data, can be tuned by using automatic tuning algorithms such as the Bayesian optimization \cite{wu2019hyperparameter}. In this way, the optimum hyperparameters for the COVID-19 detection problem can be tuned for both CNNs and SVMs to obtain higher performance.

\end{document}